\documentclass[aps,preprint,epsfig,rotate]{revtex4}
\usepackage{graphicx}
\usepackage{bm}
\usepackage{epsfig}
\begin{document}
\title{Mass-dependencies of the bound state properties for three-body positronium-like exitonic complexes}

\author{Alexei M. Frolov}
 \email[E--mail address: ]{alex1975frol@gmail.com} 

%\affiliation{ITAMP, Harvard-Smithonian Center for Astrophysics, \\
%         MS 14, 60 Garden Street, Cambridge MA 02138-1516, USA}  

\affiliation{Department of Applied Mathematics \\
 University of Western Ontario, London, Ontario N6H 5B7, Canada}

%\date{January 11, 2018}
\date{\today}

\begin{abstract}

Mass-dependencies of a number of bound state properties are investigated in some light two-electron exitonic complexes (or clusters) 
$Z^{+} e^{-} e^{-}$, where $m_e \le m_Z \le 2 m_e$. These exitonic complexes (or model ions) play a great role in modern solid state 
physics, since such complexes describe optical absorption in a number of semiconductors.  We also derived and tested a number of 
accurate mass-interpolation formulas for these properties. In general, our mass-interpolation formulas allow one to predict (fast 
and accurately) numerical values of these bound state properties in the `new' exitonic complexes, i.e., in three-body exitonic 
complexes with new mass ratios.   

\noindent 
PACS number(s): 71.35.-y and 36.10.-k

\end{abstract}

\maketitle

%\noindent \vspace{0.2in}

\section{Introduction}

In this communication we investigate the bound state properties of some two-electron clusters (or exitons) $Z^{+} e^{-} e^{-}$ with unit electrical 
charges. The notation $Z^{+} e^{-} e^{-}$ is used everywhere below in this study and designates the three-body ion with the unit electric charges, 
where the mass of this `heavy' (or central) particle $Z^{+}$ is slightly larger than unity. In fact, below we shall assume that the mass $m_Z (= M)$
of the positively charged, central particle $Z^{+}$, is bounded between $m_e$ and $ 2m_e$, i.e., $m_e \le m_Z \le 2 m_e$, or $1 \le m_Z \le 2$ in 
atomic units, where $\hbar = 1, \mid e \mid = 1$ and $m_e = 1$. In this study such three-body $Z^{+} e^{-} e^{-}$ systems (or ions) are called the 
positronium-like three-body exitonic complexes, or exitonic complexes, for short. In general, the three-body complexes ($Z^{+} e^{-} e^{-}$) are of 
great interest in modern solid state physics (see, e.g., \cite{Dav} - \cite{Thil}), since they substantially determine the actual optical absorption 
(or light absorption) in a number of important semiconductors. Stability of the two-body Coulomb exitons, such as $e^{+} e^{-}$ and $A^{+} e^{-}$, 
in real solids was predicted by Wannier \cite{Wan} and Mott \cite{Mott} in the end of 1930's. Originally, these two-body clusters, e.g., $A^{+} 
e^{-}$ and $e^{+} e^{-}$, were used to explain optical spectra and a number of properties of many semiconductors, 
including CdS, Cu$_2$O, PbI$_2$, etc. 

Later, Lampert \cite{Lamp} suggested that some three-, four- and similar few-body exitons (or exitonic complexes) can also be stable and, therefore, 
useful in applications to actual problems known in physics of semiconductors. He also introduced the system of convenient notations for such 
few-body exitons. For instance, the positively charged hole (with the unit electric charge) was designated as $h$ (or, $h^{+}$), while the negatively 
charged `effective' electron (or quasi-electron) was denoted by the letter $e$ (or $e^{-}$). In general, the effective masses of holes and electrons 
can differ significantly from their usual values. In earlier studies \cite{Dav} - \cite{Thil} the effective electron masses for light exitons were 
varied between $\approx$ 0.068 $m_e$ and $\approx$ 1.25 $m_e$, while analogous masses for the positively charged holes were varied between $\approx$ 
0.250 $m_e$ and $\approx$ 3.275 $m_e$. It is clear that all essential properties of any Coulomb three-body system with unit electrical charges 
crucially depend upon the ratio of particle masses, i.e., upon the dimensionless mass ratio $\frac{m_h}{m_e}$. At the same time, the absolute values 
of these masses play a relatively minor role for prediction of such properties.   

Note that Lampert \cite{Lamp} defined and considered two different classes of few-body exitonic complexes: (a) light exitonic comlexes, where the 
mass of the positively charged hole is comparable with the mass of electron, and (b) heavy exitonic complexes, where the mass of the central hole 
substantially exceeds the electron mass, e.g., $m_h \ge$ 1500 $m_e$. Below, we restrict ourselves to the consideration of the light (or mobile) 
exitonic complexes only. Each of these three-body systems consists of two electrons and one positively charged hole which has the mass comparable 
with unity (in $a.u.$). These light two-electron exitonic complexes (or light exitonic clusters) are of interest for future development of physics 
of semiconductors and, in particular, for optical and infrared spectroscopy of semiconductors. Indeed, the infrared and optical spectra of many 
semicondutors can now be measured accurately and with a large number of tiny details. Theoretical studies of various three- and few-body exitonic 
complexes are needed to explain all these details. As mentioned above, in this study we discuss the class of light (or mobile) exitonic complexes, 
where the mass of positively charged `hole' $m_h (= m_Z)$ is comparable with the electron mass $m_e$. This can be written in the form $m_Z = \lambda 
m_e$, where the numerical parameter $\lambda$ is bounded between 1 and 2. It is clear that such three-body ions are similar, e.g., by their structure 
and bound state properties to the negatively charged positronium ion P$s^{-}$ (or $e^{-} e^{+} e^{-}$). Briefly, this means that in our study we 
consider some positronium-like excitonic complexes, or ions, for short. As follows from the basic principles of Quantum Mechanics and famous 
Poincare theorem (see below) all bound state properties of such systems must be `smooth' functions of the same parameter $\lambda$ which is the 
dimensionless mass ratio of the heavy and light particles. Further investigations of this problem lead us to unambiguous conclusion that all regular 
bound state properties of the $Z^{+} e^{-} e^{-}$ ions must be analytical functions of the same dimensionless parameter $\lambda$ \cite{Chew}. 
Investigation of these functions and their approximation by smooth interpolation formulas is the main goal of this study.
       
\section{Hamiltonian of the three-body exitonic complexes}

In the lowest-order relativistic (or non-relativistic) approximation the Hamiltonian of an arbitrary three-body exitonic $Z^{+} e^{-} e^{-}$ complex 
(or ion) is written in the form 
\begin{eqnarray}
 H = -\frac{\hbar^2}{2 m_e} \Bigl[ \nabla^{2}_{1} + \nabla^{2}_{2} + \Bigl(\frac{m_e}{m_Z}\Bigr) \nabla^{2}_{Z} \Bigr] - \frac{e^2}{r_{32}} 
- \frac{e^2}{r_{31}} + \frac{e^2}{r_{21}} \; \; \; , \; \; \label{Hamil}
\end{eqnarray}
where $\hbar = \frac{h}{2 \pi}$ is the reduced Planck constant (or Dirac constant), $m_e$ is the electron mass, $m_1 = m_e, m_2 = m_e$ and $m_3 
= m_Z$ are the masses of three point particles and $e$ is the electric charge of electron. Here and everywhere below in this study the notation 
$e^{-}$ stands for the electron, while $Z^{+}$ (or $h$) means the positively charged hole. In atomic units (or $a.u.$, for short) the 
Hamiltonian $H$, Eq.(\ref{Hamil}), takes the form 
\begin{eqnarray}
 H = -\frac12 \Bigl[ \nabla^{2}_{1} + \nabla^{2}_{2} + \frac{1}{m_Z} \nabla^{2}_{Z} \Bigr] &-& \frac{1}{r_{32}} - \frac{1}{r_{31}} + 
 \frac{1}{r_{21}} = -\frac12 \Bigl[ \nabla^{2}_{1} + \nabla^{2}_{2} + \frac{1}{\lambda} \nabla^{2}_{Z} \Bigr] \nonumber \\
 &-& \frac{1}{r_{32}} - \frac{1}{r_{31}} + \frac{1}{r_{21}} \; \; \; , \; \; \label{Hamil1}
\end{eqnarray}
where the mass $m_Z (= m_h)$ must be expressed in terms of the electron mass $m_e$, i.e., $m_Z = \lambda m_e = \lambda$. In general, atomic units 
are very convenient to analyze all atomic, quasi-atomic and molecular systems, i.e., few-body systems with the Coulomb interaction between each 
pair of particles. In particular, these units are applied everywhere in this study, including all equations below and in all Tables.  

Now, the main goal of our current investigation can be formulated as follows. In general, by solving the non-relativistic Schr\"{o}dinger equation 
for the bound state spectrum in the $Z^{+} e^{-} e^{-}$ ions with the Coulomb Hamiltonian $H$, Eq.(\ref{Hamil1}), one finds the corresponding 
three-body wave function(s) $ \Psi({\bf r}_{1}, {\bf r}_{2}, {\bf r}_{3}, {\bf s}_{1}, {\bf s}_{2}, {\bf s}_{3}; \lambda)$ 
\begin{eqnarray}
 H({\bf r}_{1}, {\bf r}_{2}, {\bf r}_{3}; \lambda) \Psi({\bf r}_{1}, {\bf r}_{2}, {\bf r}_{3}, {\bf s}_{1}, {\bf s}_{2}, {\bf s}_{3}; \lambda) = 
 E(\lambda)  \Psi({\bf r}_{1}, {\bf r}_{2}, {\bf r}_{3}, {\bf s}_{1}, {\bf s}_{2}, {\bf s}_{3}; \lambda) \; \; \; \label{Schro}
\end{eqnarray}
where $\lambda$ is a parameter, while ${\bf r}_{1}, {\bf r}_{2}, {\bf r}_{3}$ and ${\bf s}_{1}, {\bf s}_{2}, {\bf s}_{3}$ are the Cartesian and 
spin coordinates, respectively, of the three point particles in the Coulomb three-body system $Z^{+} e^{-} e^{-}$ with unit electrical charges. 
Since all coefficients in the Hamiltonian, Eq.(\ref{Hamil1}), are the smooth (or analytical) functions of the dimensionless parameter $\lambda$, 
or its inverse parameter $\mu$, where $\mu = \frac{1}{\lambda} = \frac{m_e}{m_Z} = \frac{1}{m_Z}$, then the actual wave function $\Psi({\bf 
r}_{1}, {\bf r}_{2}, {\bf r}_{3}, {\bf s}_{1}, {\bf s}_{2}, {\bf s}_{3}; \lambda)$, Eq.(\ref{Schro}), is also a smooth function of the same 
parameter $\lambda$ (or $\mu = \frac{1}{\lambda})$. This fact follows from the well-known Poincare theorem (see, e.g., \cite{Mig1}) which states 
that there is a very close and uniform relation between singularities of the solution(s) of an arbitrary differential equation (e.g., the original 
Schr\"{o}dinger equation, Eq.(\ref{Schro})) and singularities of the coefficients in this equation (see, e.g., \cite{Mig1}). In other words, if 
there is no singularities (upon some parameters) in the coefficients of the Schr\"{o}dinger equation, then the corresponding wave functions are 
regular functions of the same parameters. From here one finds that the expectation values $\langle \Psi \mid A \mid \Psi \rangle$, where $A$ is 
some non-singular operator, computed in atomic units with the wave functions $\Psi({\bf r}_{1}, {\bf r}_{2}, {\bf r}_{3}, {\bf s}_{1}, {\bf s}_{2}, 
{\bf s}_{3}; \lambda)$, Eq.(\ref{Schro}), are also smooth (or analytical) functions of the parameter $\lambda = m_Z$, or $\mu = \frac{1}{\lambda}$ 
(in atomic units).

\section{Mass-interpolation formula}

In this study we want to investigate analytical dependencies of the different bound state properties upon the inverse mass parameter $\mu$, where 
$\mu = \frac{1}{m_Z}$. Below, we consider the whole family of light exitonic complexes $Z^{+} e^{-} e^{-}$, where the two following conditions are 
obeyed for the inverse mass $\mu = \frac{1}{m_Z}$ of the positively charged hole $h (= Z^{+})$: $\frac12 \le \mu \le 1$ (in atomic units). As 
follows from the general theory of bound states in the Coulomb three-body ions with unit charges \cite{FroBi92} for these masses $m_Z (= 
\frac{1}{\mu})$ the bound state spectrum of each of the $Z^{+} e^{-} e^{-}$ ions contains only one bound (ground) $1^{1}S(L = 0)-$state. All other 
states, including triplet states, in these two-electron ions are not bound \cite{FroBi92}.  

Let us consider the expectation value of some non-singular, spin-independent operator $B$ computed with the $\Psi({\bf r}_{1}, {\bf r}_{2}, {\bf 
r}_{3}, {\bf s}_{1}, {\bf s}_{2}, {\bf s}_{3}; \lambda)$ wave function, i.e.,  
\begin{eqnarray}
 \langle B \rangle &=& \langle \Psi({\bf r}_{1}, {\bf r}_{2}, {\bf r}_{3}, {\bf s}_{1}, {\bf s}_{2}, {\bf s}_{3}; \lambda) \mid B({\bf r}_{1}, 
 {\bf r}_{2}, {\bf r}_{3}) \mid \Psi({\bf r}_{1}, {\bf r}_{2}, {\bf r}_{3}, {\bf s}_{1}, {\bf s}_{2}, {\bf s}_{3}; \lambda) \rangle \nonumber \\
 &=& \langle \Psi \mid B \mid \Psi \rangle \; \; \; \label{prop}
\end{eqnarray}
Here and everywhere below we shall assume that the bound state wave function $\Psi$ has unit norm. As mentioned above for any non-singular 
operator $B$ its expectation value $\langle B \rangle$ is an analytical function of the parameter $\mu$, which is inverse of $\lambda$, i.e., 
$\mu = \frac{1}{\lambda}$. By using this notation and applying results from our earlier analysis of similar problem \cite{Fro2019} we can 
write the following mass-interpolation formula for the expectation value $\langle B \rangle$:
\begin{eqnarray}
 \langle B \rangle &=& \Bigl(\frac{1}{1 + 2 \mu}\Bigr) F\Bigl(\frac{1}{1 + 2 \mu}\Bigr) = \frac{B_{1}}{1 + 2 \mu} + \frac{B_{2}}{(1 + 2  
 \mu)^{2}} + \frac{B_{3}}{(1 + 2 \mu)^{3}} + \ldots = \sum^{K}_{k=1} \frac{B_{k}}{(1 + 2 \mu)^{k}}  \; \; \; \label{massA1}
\end{eqnarray}
where $\mu = \frac{1}{m_Z}$ and $B_{k}$ ($k$ = 1, 2, 3, $\ldots$) are the real coefficients in this mass-interpolation series. In this equation
the notation $\langle B \rangle$ stands for an arbitrary, in principle, bound state property (or expectation value) of the three-body $Z^{+} 
e^{-} e^{-}$ complex. For the total energies $E$ the formula, Eq.(\ref{massA1}),  was derived and tested (for similar systems) in \cite{Fro2019}. 
However, it is clear that the same formula, Eq.(\ref{massA1}), can also be used to evaluate other bound state properties in the $Z^{+} e^{-} 
e^{-}$ exitonic complexes, where $\frac12 \le \mu \le 1$ (some additional details can be found in \cite{Fro2019}).

As was found in actual computations the interpolation formula, Eq.(\ref{massA1}), is correct and accurate for any finite range of $\mu$ variation. 
However, the formula, Eq.(\ref{massA1}), is `non-physical', and, in general, all coefficients in Eq.(\ref{massA1}) are comparable to each other. 
The reason is clear, since in the formula, Eq.(\ref{massA1}), we do not have any effective small parameter. In reality, our physical intuition 
requires a different interpolation formula which contains an `obvious' small parameter(s), and such a formula should converge to the expectation 
value(s) known for some real system when this small parameter approaches to zero. In other words, the `new' mass-interpolation formula must 
produce results which can always be compared with the `exact' properties of some standard (or ethalon) system. Such `advanced' formulas do exist, 
and they can also be constructed in a variety of different ways. For instance, the following mass-interpolation formula includes one `small' 
parameter $1 - \mu = \frac{\lambda - 1}{\lambda}$, where $\lambda = m_Z$ and $0 \le 1 - \mu \le \frac12$:
\begin{eqnarray}
 \langle B \rangle &=& C_0 + C_1 \Bigl( \frac{\lambda - 1}{\lambda} \Bigr) + C_2 \Bigl( \frac{\lambda - 1}{\lambda} \Bigr)^{2} + 
 C_3 \Bigl( \frac{\lambda - 1}{\lambda} \Bigr)^{3} + \ldots = \sum^{K}_{k=0} C_k \Bigl( \frac{\lambda - 1}{\lambda} \Bigr)^{k} 
 \; \; \; \label{massA3} \\
  &=& \langle B({\rm Ps}^{-}) \rangle + \sum^{K}_{k=1} C_k \Bigl( 1 - \mu \Bigr)^{k} \nonumber 
\end{eqnarray}
where $C_0 = \langle B({\rm Ps}^{-}) \rangle$, i.e., if $1 - \mu = 0$ (or $\lambda = 1$ and $\mu = 1$), then the expectation value $\langle B 
\rangle$ coincides with the numerical value, which was determined in the direct numerical computations of the ground $1^{1}S(L = 0)-$state of 
the Ps$^{-}$ ion. This interpolation formula is applied below to represent our results obtained for three-body exitonic complexes $Z^{+} e^{-} 
e^{-}$. Note that this formula, Eq.(\ref{massA3}), is a power expansion upon the positive powers of the small parameter $1 - \mu = \frac{\lambda 
- 1}{\lambda}$, where $\lambda  = m_Z$ for three-body clusters (or exitons) $Z^{+} e^{-} e^{-}$. The derived mass-interpolation formula, 
Eq.(\ref{massA3}), is accurate and numerically stable, if the dimensionless parameter $\lambda$ varies between 1 and 2.  

\section{Exponential Variational Expansion}

To perform highly-accurate computations of the bound $1^{1}S(L = 0)-$states in the $Z^{+} e^{-} e^{-}$ ions we apply our exponential variational 
expansion \cite{Fro2015} in the relative coordinates $r_{32}, r_{31}$ and $r_{21}$, where $r_{ij} = \mid {\bf r}_i - {\bf r}_j \mid = r_{ji}$ and 
$(i,j,k)$ = (1,2,3). In fact, for these three-body systems we also used the exponential variational expansion written in the perimetric three-body 
coordinates $u_1, u_2$ and $u_3$ \cite{Fro2015}, which provide much better overall accuracy for the bound state computations (see below). First, 
let us consider three-body exponential variational expansion in the relative coordinates $r_{32}, r_{31}$ and $r_{21}$, where $r_{ij} = \mid {\bf 
r}_i - {\bf r}_j \mid = r_{ji}$ are the three scalar interparticle distances. For the bound ${}^{1}S(L = 0)-$states in the $Z^{+} e^{-} e^{-}$ ions 
this expansion takes the form (see, e.g., \cite{Fro2015}) 
\begin{eqnarray}
 \Psi(r_{32}, r_{31}, r_{21}) = \frac12 \Bigl( 1 + \hat{P}_{21} \Bigr) \sum_{i=1}^{N} C_{i} \exp(-\alpha_{i} r_{32} - \beta_{i} r_{31} - 
 \gamma_{i} r_{21}) \; , \; \label{equ53} 
\end{eqnarray}
where $\alpha_{i}, \beta_{i}, \gamma_{i}$ ($i = 1, \ldots, N$) are the $3 N$ non-linear parameters and $N$ is the total number of basis functions, 
i.e., the total number of exponents in Eq.(\ref{equ53}). The notation $\hat{P}_{21}$ in Eq.(\ref{equ53}) stands for the operator which produces 
permutation of the two identical particles (two electrons) in the light three body exitonic complex $Z^{+} e^{-} e^{-}$ (or $h e e$). The 
exponential variational expansion, Eq.(\ref{equ53}), has many advantages in actual bound state computations of various three-body systems, e.g., 
it has an incredible flexibility, and, it also provides a very high accuracy for the final results. Another group of advantages includes an obvous 
simplicity of all expressions arising for the matrix elements of the Hamiltonian and other operators which are needed in calculations of various 
bound state properties. This directly follows from the well known simplicity of original formulas for the Laplace transformations of these 
operators. Numerous computational advantages of Eq.(\ref{equ53}) are related to large numbers of non-linear parameters $\alpha_{i}, \beta_{i}, 
\gamma_{i}$, where $i = 1, \ldots, N$, which are carefully optimized in computations. However, an obvious disadvantage of the variational 
expansion, Eq.(\ref{equ53}), follows from the fact that the three relative coordinates $r_{32}, r_{31}$ and $r_{21}$ are not independent from 
each other. Indeed, the six following inequalities $\mid r_{ik} - r_{jk} \mid \le r_{ij} \le r_{ik} + r_{jk}$ are always obeyed for the three 
relative coordinates, where $(i,j,k)$ = (1,2,3). These aditional conditions substantially complicate actual optimization of the non-linear 
parameters in Eq.(\ref{equ53}). 

Fortunately, for three-body systems there is another (or alternative) set of the three perimetric coordinates $u_1, u_2, u_3$ which are truly 
independent of each other and each of them varies between 0 and $+\infty$. These perimetric coordinates have been introduced in physics of 
three-body systems by C.L. Pekeris in \cite{Pek1}. The three perimetric coordinates are simply (even linearly) related to the relative 
coordinates and vice versa 
\begin{eqnarray}
  & & u_1 = \frac12 ( r_{21} + r_{31} - r_{32}) \; \; \; , \; \; \; r_{32} = u_2 + u_3 \nonumber \\
  & & u_2 = \frac12 ( r_{21} + r_{32} - r_{31}) \; \; \; , \; \; \; r_{31} = u_1 + u_3 \; \; \; \label{coord} \\
  & & u_3 = \frac12 ( r_{31} + r_{32} - r_{21}) \; \; \; , \; \; \; r_{21} = u_1 + u_2 \nonumber
\end{eqnarray}
where $r_{ij} = r_{ji}$. For the bound ${}^{1}S(L = 0)-$states our exponential variational expansion in these three-body perimetric coordinates 
takes the form (more details are discussed, e.g., in \cite{Fro2015}):
\begin{eqnarray}
 \Psi(u_1, u_2, u_3) = \frac12 \Bigl( 1 + \hat{P}_{21} \Bigr) \sum_{i=1}^{N} C_{i} \exp(-\alpha_{i} u_1 - \beta_{i} u_2 - \gamma_{i} u_3)  
 \label{equ55} 
\end{eqnarray}
where $u_{1}, u_{2}, u_{3}$ are the three perimetric coordinates and all $3 N-$non-linear parameters $\alpha_{i}, \beta_{i}, \gamma_i$ ($i = 1, 
\ldots, N$) in Eq.(\ref{equ55}) are real, independent of each other and always positive. During optimization of these parameters we do not need 
to check any additional condition for these parameters (compare with Eq.(\ref{equ53})). In other words, the true independence of perimetric 
coordinates $u_1, u_2, u_3$ allows one to simplify drastically the whole process of optimization of the non-linear parameters in Eq.(\ref{equ55}). 

Note that there are three additional conditions for the non-linear parameters which follow from the completness of the radial set of exponential 
basis functions, Eq.(\ref{equ55}), \cite{XX}, but these conditions have nothing to do with the optimization of the non-linear 
parameters. Indeed, one can show that the set of exponential `radial' functions is complete, if (and only if) the three following series of 
inverse powers of the non-linear parameters in Eq.(\ref{equ55}) are divergent, i.e., the three following sums (or series): $S_1 = \sum_{i=1} 
\frac{1}{\alpha_{i}}, S_2 = \sum_{i=1} \frac{1}{\beta_{i}}$ and $S_3 = \sum_{i=1} \frac{1}{\gamma_{i}}$ are divergent (or become infinite) 
when $i \rightarrow \infty$. As follows from here, these three conditions do not complicate optimization of the non-linear parameters in 
Eq.(\ref{equ55}).    

To increase the overall efficiency of our exponential variational expansion, Eq.(\ref{equ55}), even further, in actual numerical calculations we 
applied the two-stage optimization procedure of the non-linear parameters \cite{Fro1998}. At the first stage of this procedure one needs to 
construct the first-stage short-term wave function which is very accurate and usually contains $N \approx$ 400 - 800 of basis functions, or 
exponents, in Eq.(\ref{equ55}). All non-linear parameters in this short-term wave function (or first-stage cluster) are carefully optimized to 
provide a very high accuracy. At the second stage we construct the optimal tail (or `orthogonal complement') of the short-term wave function 
constructed at the first stage. In calculations performed for this study the `optimized' tail of our variational wave function contained 
somewhere between 3500 and 4500 basis functions, i.e., exponents in perimetric coordinates. Optimization of the non-linear parameters in the 
short-term cluster and in the tail of the complete (trial) wave function is described in detail in our papers \cite{Fro2001}. 

\section{Results and discussions}  

Results of our numerical computations of the ground (bound) $1^{1}S(L = 0)-$states in the three-body exitonic $Z^{+} e^{-} e^{-}$ complexes with 
unit charges can be found in Tables I - VII which contain our results for twenty two different (light) exitonic complexes $Z^{+} e^{-} e^{-}$, 
where $m_e \le m_Z \le 2 m_e$. All bound state properties presented in these Tables are expressed in atomic units ($a.u.$). Table I includes the 
total energies of these light exitonic complexes $Z^{+} e^{-} e^{-}$. Tables II - V contain a number of bound state properties determined for 
each ion considered in this study. Table VI represents the list of bound state properties of the exitonic complex $(1.75)^{+} e^{-} e^{-}$ which 
are routinely determined in modern highly accurate computations of three-body systems. Physical meaning of all notations used to designate these 
bound state properties in Table VI is explained in detail in \cite{Fro2015}. Each of these bound state properties shown in Table VI can also be 
determined for other exitonic complexes $Z^{+} e^{-} e^{-}$ mentioned in this study. Then we can construct analogous mass-interpolation formulas 
for these `additional' bound state properties. Table VII contains the coefficients $C_k$ in the mass-interpolation formula, Eq.(\ref{massA3}), 
determined for the bound state properties of the two-electron $Z^{+} e^{-} e^{-}$ exitonic complexes presented in Tables I - V. 

In general, if the coefficients $C_i$ ($i = 1, 2, \ldots, N$) in the formula, Eq.(\ref{massA3}), for some bound state property of the three-body 
$Z^{+} e^{-} e^{-}$ ions are known, then we can predict (to relatively high accuracy) the numerical value of the same property in the new 
three-body $Y^{+} e^{-} e^{-}$ ion(s) which has a different mass ratio $\mu$. In other words, by using our current results one can predict the 
bound state properties of new systems. For three-body exitonic complexes with unit charges this problem is easily reversed: we can predict the 
mass ratios in these systems which provide $a$ $priory$ given numerical values for some bound state properties. Note that the actual convergence 
rates are usually high (even very high) for many bound state properties of the three-body $Z^{+} e^{-} e^{-}$ ions, where the mass of 
$Z^{+}$-particle, i.e., $m_Z$, is close to unity. The overall accuracy of our mass-interpolation formula, Eq.(\ref{massA3}), is also very high for 
all such properties. In particular, this is true for the $\langle r^{k}_{ij} \rangle$ expectation values, where $k = -1, 1, 2$, for all expectation 
values of the interparticle $cosine-$functions, for  single particle momenta $\langle \frac12 p^{2}_{i} \rangle$ and some other expectation values, 
the observed accuracy of our interpolation formula, Eq.(\ref{massA3}), is very high. Furthermore, for these properties such an accuracy increases 
when the total accuracy of the wave function also increases.  

However, there are some bound state properties in such three-body ions for which the last statement is not always true. For instance, one finds 
some problems with the expectation values of all delta-functions in three-body systems, i.e., for the $\langle \delta({\bf r}_{+-}) \rangle, 
\langle \delta({\bf r}_{--}) \rangle$ and $\langle \delta({\bf r}_{+--}) \rangle$ delta-functions. Accurate numerical computations of the triple 
delta-function $\langle \delta({\bf r}_{+--}) \rangle$ is very difficult, since the formulas for matrix elements of the $\delta({\bf r}_{+--})$ 
operator do not contain any of the non-linear parameter from our exponential expansion, Eq.(\ref{equ55}). In other words, our variational 
expansion, Eq.(\ref{equ55}), is not flexible enough to determine the $\langle \delta({\bf r}_{+--}) \rangle$ expectation value to high accuracy. 
Similar problems arise for the so-called singular expectation values which are needed to determine some important properties in many actual 
three-body ions and atoms. For instance, these expectation values are used to determine the lower-order relativistic and QED-corrections in 
three-body systems. The general theory of singular expectation values for three-body systems is very complex and even classification of all 
possible singular integrals is not completed even for three-body systems. However, the technique of analytical and numerical computations of some 
special singular expectation values, e.g., the power-type expectation values, are well developed (see, e.g., \cite{HFS}, \cite{Fro2005} and 
references therein). It can be shown that such singular, power-type expectation values always include either delta-functions, or products of these 
delta-functions with the partial derivatives of different orders in respect to the relative coordinates $r_{ij}$ (see, e.g., formulas (28), (43) 
and (54) in \cite{Fro2005}). A different group of problems can be found so-called slow convergent bound state properties, e.g., for the $\langle 
r^{k}_{ij} \rangle$ expectation values, where $k \ge 6$. For these properties our mass-interpolation formula, Eq.(\ref{massA3}), does not provide 
very high numerical accuracy. The problems related to accurate computations of similar, i.e., singular and slow convergent, expectation values 
will be considered elsewhere. 

\section{Conclusion}

We have studied a number of bound state properties of the ground $1^{1}S(L = 0)-$states in some light three-body (or two-electron) exitonic 
complexes $Z^{+} e^{-} e^{-}$ (or $Z^{+} e^{-}_{2}$) with the unit electric charges. For these properties we have considered their 
mass-dependencies, i.e., dependencies of the corresponding expectation values upon the mass ratio of light ($e^{-}$) and heavy ($Z^{+}$) 
particles. The bound state properties discussed in this study are analytical (or smooth) functions upon particle masses, or upon their mass 
ratio(s), e.g., $\lambda = \frac{m_Z}{m_e}$, or $\mu = \frac{m_e}{m_Z}$. Finally, we derived and tested a number of accurate 
mass-interpolation formulas which describe (to very high accuracy) the actual mass-dependencies ofthese properties. It was also found in 
real application that our mass-interpolation formulas allow one to predict (fast and accurately) numerical values of these bound state 
properties for the `new' exitonic complexes, i.e., for three-body exitonic complexes with new mass ratios. By using this approach one can 
create some new three-, four- and few-body exitonic complexes with the $a$ $priory$ known bound state properties. This means that by varying 
particle masses in similar exitonic complexes we can change optical absorption in a number of semiconductors. In this form such a problem was 
not considered in earlier studies. 

Note that a number of mass-asymptotic and mass-interpolation formulas were used to approximate the total energies of the ground (bound) states 
in different three-, four and few-body systems. In very few cases analogous formulas were used for the total energies of some excited states in 
Coulomb few-body systems. However, similar mass-asymptotic and mass-interpolation formulas have never been used to interpolate (or extrapolate) 
numerical values of other bound state properties in different few-body systems. In general, for Coulomb few-body systems numerical methods based 
on the actual mass-interpolation are practically unknown. The procedure itself was not developed in earlier studies. In contrast with this, 
analogous procedures based on interpolation of the total energy and other bound state properties upon the ratio of electron and nuclear electric 
charges, i.e., upon the $\frac{e}{Q e} = \frac{1}{Q}$ ratio. The corresponding $Q^{-1}$-series are often called the $Q^{-1}-$expansions. This 
technique works very well in applications to iso-electronic series, which include positively charged ions and neutral atoms with the same number 
of bound electrons (see, e.g.,  \cite{Eps}, \cite{BS} and references therein). General theory of these $Q^{-1}-$expansions is well developed 
\cite{Shull} - \cite{Fro15}. In particular, currently we know analytical expressions for all coefficients in these formulas (see, e.g., 
\cite{Fro15}) for an arbitrary $N_e-$electron ion/atom with a given nuclear charge $Q$, where $Q \ge N_e$ (the case of negatively charged atomic 
ions is fundamentally different). For the mass-interpolation formulas we still do not have analogous transparent and logically closed theory. 
Therefore, to determine numerical coefficients in actual mass-interpolation formulas one needs to apply data from highly accurate computations.

\newpage
\begin{table}[tbp]
\caption{The total energies $E$ (in atomic units $a.u.$) of the ground $1^{1}S-$states of the two-electron $Z^{+} e^{-} e^{-}$ exitonic complexes.
         The masses of the $Z^{+}-$particle is indicated by the parameter $\lambda$, i.e., $m_Z = \lambda m_e$. For the Ps$^{-}$ ion this parameter 
         ($\lambda$) equals unity.}
     \begin{center}
     \scalebox{0.85}{%
     \begin{tabular}{| l | l | l | l |}
         \hline\hline
 system  & $E$ & system & $E$ \\  
     \hline
 (Ps$^{-})^{(a)}$              & -0.26200507 02329801 07770400 325   & $(1.30 m_e)^{+} e^{-} e^{-}$ & -0.29566085 99837913 82219162 \\        
 $(1.05 m_e)^{+} e^{-} e^{-}$  & -0.26828833 88776398 29145236       & $(1.35 m_e)^{+} e^{-} e^{-}$ & -0.30044700 03102747 35045868 \\      
 $(1.10 m_e)^{+} e^{-} e^{-}$  & -0.27427631 41008049 37604805       & $(1.40 m_e)^{+} e^{-} e^{-}$ & -0.30503705 43789494 01431516 \\     
 $(1.15 m_e)^{+} e^{-} e^{-}$  & -0.27998964 19503118 59367952       & $(1.45 m_e)^{+} e^{-} e^{-}$ & -0.30944296 86282374 83627970 \\ 
 $(1.20 m_e)^{+} e^{-} e^{-}$  & -0.28544706 01274000 74064702       & $(1.50 m_e)^{+} e^{-} e^{-}$ & -0.31367572 83583423 20973190 \\                                       
 $(1.25 m_e)^{+} e^{-} e^{-}$  & -0.29066561 68938227 79488962       & $(1.55 m_e)^{+} e^{-} e^{-}$ & -0.31774545 34647696 72372974 \\      
        \hline\hline 
 $(1.60 m_e)^{+} e^{-} e^{-}$  & -0.32166148 28290255 23311967     & $(1.85 m_e)^{+} e^{-} e^{-}$ & -0.33921711 05014934 48444183 \\ 
 $(1.65 m_e)^{+} e^{-} e^{-}$  & -0.32543244 89231741 53982378     & $(1.90 m_e)^{+} e^{-} e^{-}$ & -0.34237176 55866827 72681558 \\   
 $(1.70 m_e)^{+} e^{-} e^{-}$  & -0.32906634 39420381 29045744     & $(1.95 m_e)^{+} e^{-} e^{-}$ & -0.34542155 52607940 50089018 \\   
 $(1.75 m_e)^{+} e^{-} e^{-}$  & -0.33257057 85764355 81210798     & $(1.99 m_e)^{+} e^{-} e^{-}$ & -0.34778938 07581639 64154601 \\   
 $(1.80 m_e)^{+} e^{-} e^{-}$  & -0.33595203 43747174 01041876     & $(2.00 m_e)^{+} e^{-} e^{-}$ & -0.34837166 58903586 37841268 \\ 
       \hline\hline 
  \end{tabular}}
  \end{center}
  ${}^{(a)}$Or $(m_e)^{+} e^{-} e^{-}$ in our current notations.
  \end{table}
\newpage
\begin{table}[tbp]
\caption{The mass-dependencies (or $\lambda$-dependencies, where $\lambda = \frac{m_Z}{m_e} = m_Z$) of some powers of electron-hole distances 
         (in atomic units $a.u.$) in the ground $1^{1}S-$states of the two-electron $Z^{+} e^{-} e^{-}$ exitonic complexes.}
     \begin{center}
     \scalebox{0.85}{%
     \begin{tabular}{| l | c | c | c |}
       \hline\hline
 system  & $\langle r^{-1}_{+-} \rangle$ & $\langle r_{+-} \rangle$ & $\langle r^{+2}_{+-} \rangle$ \\  
     \hline
 (Ps$^{-})^{(a)}$              & 0.339821023059220307 & 5.489633252359449933 & 48.4189372262379554 \\    
 $(1.05 m_e)^{+} e^{-} e^{-}$  & 0.347751777477306284 & 5.377410993019180797 & 46.5954306785745991 \\    
 $(1.10 m_e)^{+} e^{-} e^{-}$  & 0.355311776963369283 & 5.274457662477727990 & 44.9475366106200294 \\
 $(1.15 m_e)^{+} e^{-} e^{-}$  & 0.362527372402817319 & 5.179620975891703988 & 43.4506353144262322 \\  
 $(1.20 m_e)^{+} e^{-} e^{-}$  & 0.369422386517961194 & 5.091936574020034224 & 42.0845435291501284 \\                                      
 $(1.25 m_e)^{+} e^{-} e^{-}$  & 0.376018419877168233 & 5.010590800115592042 & 40.8325567000539007 \\
                     \hline\hline        
 $(1.30 m_e)^{+} e^{-} e^{-}$  & 0.382335112226246589 & 4.934892022211731321 & 39.6807284822216131 \\
 $(1.35 m_e)^{+} e^{-} e^{-}$  & 0.388390366762103012 & 4.864248294237640165 & 38.6173219083531709 \\
 $(1.40 m_e)^{+} e^{-} e^{-}$  & 0.394200543485897923 & 4.798149777321126420 & 37.6323864882014287 \\
 $(1.45 m_e)^{+} e^{-} e^{-}$  & 0.399780626613580077 & 4.736154776507943084 & 36.7174288679186219 \\
 $(1.50 m_e)^{+} e^{-} e^{-}$  & 0.405144370107756600 & 4.677878552349606113 & 35.8651538171963564 \\
 $(1.55 m_e)^{+} e^{-} e^{-}$  & 0.410304424668904616 & 4.622984283036264619 & 35.0692586576426834 \\        
                \hline\hline 
 $(1.60 m_e)^{+} e^{-} e^{-}$  & 0.415272448943228632 & 4.571175708369267399 & 34.3242687123735936 \\
 $(1.65 m_e)^{+} e^{-} e^{-}$  & 0.420059207236906119 & 4.522191100177443248 & 33.6254045409954058 \\  
 $(1.70 m_e)^{+} e^{-} e^{-}$  & 0.424674655647693447 & 4.475798287182586144 & 32.9684740212292575 \\  
 $(1.75 m_e)^{+} e^{-} e^{-}$  & 0.429128018216262981 & 4.431790524331000775 & 32.3497840139571345 \\   
 $(1.80 m_e)^{+} e^{-} e^{-}$  & 0.433427854446854387 & 4.389983043153442103 & 31.7660675833384012 \\
                 \hline\hline 
 $(1.85 m_e)^{+} e^{-} e^{-}$  & 0.437582119338701449 & 4.350210154964120217 & 31.2144236626217027 \\
 $(1.90 m_e)^{+} e^{-} e^{-}$  & 0.441598216897524429 & 4.312322805626327622 & 30.6922667464344848 \\
 $(1.95 m_e)^{+} e^{-} e^{-}$  & 0.445483047953298614 & 4.276186501329255722 & 30.1972847131176064 \\
 $(1.99 m_e)^{+} e^{-} e^{-}$  & 0.448500743386934620 & 4.248456037074223472 & 29.8194638856757956 \\
 $(2.00 m_e)^{+} e^{-} e^{-}$  & 0.449243052991092990 & 4.241679540884605032 & 29.7274032798995683 \\
                     \hline\hline        
  \end{tabular}}
  \end{center}
  ${}^{(a)}$Or $(m_e)^{+} e^{-} e^{-}$ in our current notations. 
  \end{table}
\newpage
\begin{table}[tbp]
\caption{The mass-dependencies (or $\lambda$-dependencies, where $\lambda = \frac{m_Z}{m_e} = m_Z$) of some powers of electron-electron distances 
         (in atomic units $a.u.$) in the ground $1^{1}S-$states of the two-electron $Z^{+} e^{-} e^{-}$ exitonic complexes.}
     \begin{center}
     \scalebox{0.85}{%
     \begin{tabular}{| l | c | c | c |}
       \hline\hline
 system  & $\langle r^{-1}_{--} \rangle$ & $\langle r_{--} \rangle$ & $\langle r^{2}_{--} \rangle$ \\  
     \hline
 (Ps$^{-})^{(a)}$              & 0.1556319056524803974 & 8.548580655099186112 & 93.17863384798132901 \\       
 $(1.05 m_e)^{+} e^{-} e^{-}$  & 0.1589268771993329095 & 8.389716655318915275 & 89.91111200998956544 \\  
 $(1.10 m_e)^{+} e^{-} e^{-}$  & 0.1620709257251286909 & 8.243505167923295249 & 86.94856383522881360 \\   
 $(1.15 m_e)^{+} e^{-} e^{-}$  & 0.1650754609050109189 & 8.108405664314476662 & 84.24922876288355002 \\
 $(1.20 m_e)^{+} e^{-} e^{-}$  & 0.1679506527811222393 & 7.983126334507210675 & 81.77874302697318085 \\                                     
 $(1.25 m_e)^{+} e^{-} e^{-}$  & 0.1707056059666909060 & 7.866575051392824220 & 79.50856336949266407 \\     
                     \hline\hline        
 $(1.30 m_e)^{+} e^{-} e^{-}$  & 0.1733485044849104128 & 7.757821564196609950 & 77.41477722241752222 \\
 $(1.35 m_e)^{+} e^{-} e^{-}$  & 0.1758867329036565529 & 7.656068027039734647 & 75.47719344793282872 \\
 $(1.40 m_e)^{+} e^{-} e^{-}$  & 0.1783269782138970425 & 7.560625790557808915 & 73.67863953916683900 \\
 $(1.45 m_e)^{+} e^{-} e^{-}$  & 0.1806753159706851857 & 7.470896953022572330 & 72.00441266850635735 \\
 $(1.50 m_e)^{+} e^{-} e^{-}$  & 0.1829372834988285580 & 7.386359566343292185 & 70.44184671331587430 \\
 $(1.55 m_e)^{+} e^{-} e^{-}$  & 0.1851179424082698875 & 7.306555676058228247 & 68.97996765492341739 \\        
                \hline\hline 
 $(1.60 m_e)^{+} e^{-} e^{-}$  & 0.1872219322284062164 & 7.231081578756334237 & 67.60921699320567703 \\
 $(1.65 m_e)^{+} e^{-} e^{-}$  & 0.1892535166274639301 & 7.159579829227156132 & 66.32122799903554617 \\  
 $(1.70 m_e)^{+} e^{-} e^{-}$  & 0.1912166234113106365 & 7.091732639255718633 & 65.10864337327622437 \\  
 $(1.75 m_e)^{+} e^{-} e^{-}$  & 0.1931148792796548017 & 7.027256391517860208 & 63.96496562046775744 \\   
 $(1.80 m_e)^{+} e^{-} e^{-}$  & 0.1949516401442739712 & 6.965897053256562590 & 62.88443346913977699 \\
                 \hline\hline 
 $(1.85 m_e)^{+} e^{-} e^{-}$  & 0.1967300176744160001 & 6.907426320799983055 & 61.86191918021049155 \\
 $(1.90 m_e)^{+} e^{-} e^{-}$  & 0.1984529026216833130 & 6.851638361409822936 & 60.89284272110118229 \\
 $(1.95 m_e)^{+} e^{-} e^{-}$  & 0.2001229853850091279 & 6.798347046223620670 & 59.97309964566827192 \\ 
 $(1.99 m_e)^{+} e^{-} e^{-}$  & 0.2014227252575413117 & 6.757397624161222631 & 59.27033312188854339 \\
 $(2.00 m_e)^{+} e^{-} e^{-}$  & 0.2017427742014687036 & 6.747383589208909783 & 59.09900018005077075 \\
                     \hline\hline        
  \end{tabular}}
  \end{center}
  ${}^{(a)}$Or $(m_e)^{+} e^{-} e^{-}$ in our current notations. 
  \end{table}
\newpage
\begin{table}[tbp]
\caption{The mass-dependencies (or $\lambda$-dependencies (where $\lambda = \frac{m_Z}{m_e} = m_Z$) for the electron-hole $cosine-$function 
         and some delta-functions (in atomic units $a.u.$) in the ground $1^{1}S-$states of the two-electron $Z^{+} e^{-} e^{-}$ exitonic complexes.}
     \begin{center}
     \scalebox{0.85}{%
     \begin{tabular}{| l | c | c | c |}
       \hline\hline
 system  & $\langle {\bf n}_{32} \cdot {\bf n}_{31} \rangle$ & $\langle \delta({\bf r}_{+-}) \rangle$ & $\langle \delta({\bf r}_{--}) \rangle$ \\  
     \hline
 (Ps$^{-})^{(a)}$              & 0.59198170114890223326 & 0.0207331980052196 & 0.00017099675635587 \\      
 $(1.05 m_e)^{+} e^{-} e^{-}$  & 0.59321650222396424254 & 0.0222654059625811 & 0.00018808858185933 \\
 $(1.10 m_e)^{+} e^{-} e^{-}$  & 0.59439397486701339284 & 0.0237932341957713 & 0.00020547373746425 \\
 $(1.15 m_e)^{+} e^{-} e^{-}$  & 0.59551862916851168471 & 0.0253137569372155 & 0.00022309961066071 \\ 
 $(1.20 m_e)^{+} e^{-} e^{-}$  & 0.59659444935871805741 & 0.0268244815677388 & 0.00024091878456868 \\                                      
 $(1.25 m_e)^{+} e^{-} e^{-}$  & 0.59762497478090597071 & 0.0283232947449509 & 0.00025888859872420 \\     
                     \hline\hline        
 $(1.30 m_e)^{+} e^{-} e^{-}$  & 0.59861336574741563798 & 0.0298084147633667 & 0.00027697072996527 \\
 $(1.35 m_e)^{+} e^{-} e^{-}$  & 0.59956245752560073727 & 0.0312783495504258 & 0.00029513079887424 \\ 
 $(1.40 m_e)^{+} e^{-} e^{-}$  & 0.60047480492180747513 & 0.0327318597131381 & 0.00031333800434441 \\ 
 $(1.45 m_e)^{+} e^{-} e^{-}$  & 0.60135271935771764414 & 0.0341679260808142 & 0.00033156478697562 \\
 $(1.50 m_e)^{+} e^{-} e^{-}$  & 0.60219829990605185441 & 0.0355857212305214 & 0.00034978652080524 \\ 
 $(1.55 m_e)^{+} e^{-} e^{-}$  & 0.60301345943117862625 & 0.0369845845279932 & 0.00036798123219471 \\       
                \hline\hline 
 $(1.60 m_e)^{+} e^{-} e^{-}$  & 0.6037999467361331217 & 0.03836400026186413 & 0.00038612934418938 \\ 
 $(1.65 m_e)^{+} e^{-} e^{-}$  & 0.6045593654306838275 & 0.03972357849903850 & 0.00040421344478006 \\ 
 $(1.70 m_e)^{+} e^{-} e^{-}$  & 0.6052931900908556342 & 0.04106303832163334 & 0.00042221807681767 \\   
 $(1.75 m_e)^{+} e^{-} e^{-}$  & 0.6060027801681618791 & 0.04238219315412792 & 0.00044012954990040 \\   
 $(1.80 m_e)^{+} e^{-} e^{-}$  & 0.6066893920189694021 & 0.04368093793715705 & 0.00045793575973865 \\ 
                 \hline\hline 
 $(1.85 m_e)^{+} e^{-} e^{-}$  & 0.6073541893551858825 & 0.04495923786278648 & 0.00047562605084867 \\
 $(1.90 m_e)^{+} e^{-} e^{-}$  & 0.6079982523625379754 & 0.04621711856895432 & 0.00049319105859290 \\
 $(1.95 m_e)^{+} e^{-} e^{-}$  & 0.6086225856888824136 & 0.04745465752057616 & 0.00051062259153735 \\
 $(1.99 m_e)^{+} e^{-} e^{-}$  & 0.6091084782026164451 & 0.04843012312896968 & 0.00052446688894713 \\
 $(2.00 m_e)^{+} e^{-} e^{-}$  & 0.6092281254698200062 & 0.04867197648290799 & 0.00052791351536679 \\
                     \hline\hline        
  \end{tabular}}
  \end{center}
  ${}^{(a)}$Or $(m_e)^{+} e^{-} e^{-}$ in our current notations.
  \end{table}
\newpage
\begin{table}[tbp]
\caption{The mass-dependencies (or $\lambda$-dependencies (where $\lambda = \frac{m_Z}{m_e} = m_Z$) for the partial kinetic energies and  
         factor $f = \hat{f}$ (in atomic units $a.u.$) in the ground $1^{1}S-$states of the two-electron $Z^{+} e^{-} e^{-}$ exitonic 
         complexes.}
     \begin{center}
     \scalebox{0.85}{%
     \begin{tabular}{| l | c | c | c |}
       \hline\hline 
 system  & $\langle \frac12 p^{2}_e \rangle$ & $\langle \frac12 p^{2}_h \rangle$ & $\langle \hat{f} \rangle$ \\  
     \hline
 (Ps$^{-})^{(a)}$              & 0.06661929453589000853 & 0.128766481161200091 & 0.050933258778734117068 \\   
 $(1.05 m_e)^{+} e^{-} e^{-}$  & 0.06975806633865467857 & 0.135210816510346996 & 0.050788436653496192550 \\
 $(1.10 m_e)^{+} e^{-} e^{-}$  & 0.07281700558699871945 & 0.141506533219488249 & 0.050655812455092447125 \\  
 $(1.15 m_e)^{+} e^{-} e^{-}$  & 0.07579724847235954237 & 0.147654416756431691 & 0.050534103804745326635 \\
 $(1.20 m_e)^{+} e^{-} e^{-}$  & 0.07870020624400580050 & 0.153655977167266168 & 0.050422190024038688494 \\                                      
 $(1.25 m_e)^{+} e^{-} e^{-}$  & 0.08152749447548289512 & 0.159513284928571237 & 0.050319088233439429416 \\      
                     \hline\hline        
 $(1.30 m_e)^{+} e^{-} e^{-}$  & 0.08428087662901165672 & 0.165228838743498490 & 0.050223933538545330132 \\
 $(1.35 m_e)^{+} e^{-} e^{-}$  & 0.08696221904756547255 & 0.170805458990444116 & 0.050135962501933645854 \\
 $(1.40 m_e)^{+} e^{-} e^{-}$  & 0.08957345510595255867 & 0.176246201833861998 & 0.050054499277681682437 \\
 $(1.45 m_e)^{+} e^{-} e^{-}$  & 0.09211655672017801827 & 0.181554290022428098 & 0.049978943919696348462 \\
 $(1.50 m_e)^{+} e^{-} e^{-}$  & 0.09459351178028140094 & 0.186733057196669279 & 0.049908762476578796333 \\
 $(1.55 m_e)^{+} e^{-} e^{-}$  & 0.09700630635938421199 & 0.191785903156301935 & 0.049843478563621962606 \\        
                     \hline\hline        
 $(1.60 m_e)^{+} e^{-} e^{-}$  & 0.09935691077864936465 & 0.196716258034762870 & 0.049782666162855333488 \\
 $(1.65 m_e)^{+} e^{-} e^{-}$  & 0.10164726878776087307 & 0.201527553723626473 & 0.049725943449213269035 \\
 $(1.70 m_e)^{+} e^{-} e^{-}$  & 0.10387928926368988972 & 0.206223201204919194 & 0.049672967478097130986 \\  
 $(1.75 m_e)^{+} e^{-} e^{-}$  & 0.10605483994484101561 & 0.210806572701818711 & 0.049623429599164072226 \\ 
 $(1.80 m_e)^{+} e^{-} e^{-}$  & 0.10817574280928402162 & 0.215280987761068844 & 0.049577051484840645057 \\
                     \hline\hline        
 $(1.85 m_e)^{+} e^{-} e^{-}$  & 0.11024377077941447205 & 0.219649702543929333 & 0.049533581681128074826 \\
 $(1.90 m_e)^{+} e^{-} e^{-}$  & 0.11226064549475081000 & 0.223915901734644190 & 0.049492792603723594822 \\
 $(1.95 m_e)^{+} e^{-} e^{-}$  & 0.11422803594256171570 & 0.228082692582557706 & 0.049454477915082244755 \\
 $(1.99 m_e)^{+} e^{-} e^{-}$  & 0.11576740714429035352 & 0.231346587274470682 & 0.049425481216124121014 \\           
 $(2.00 m_e)^{+} e^{-} e^{-}$  & 0.11614755777490399556 & 0.232153100681101293 & 0.049418450228367137805 \\
                     \hline\hline        
  \end{tabular}}
  \end{center}
  ${}^{(a)}$Or $(m_e)^{+} e^{-} e^{-}$ in our current notations. 
  \end{table}
\begin{table}[tbp]
   \caption{The expectation values of some propeties (in atomic units) for the $(1.75 m_e)^{+} e^{-} e^{-}$ ion. The notations `+' and/or 3 designate 
            the positively charged particle, while the notation `-', or 1 and 2 stand for electrons.}
     \begin{center}
%     \scalebox{0.72}{%
     \begin{tabular}{| c | c | c | c |}
      \hline\hline
 $\langle r^{-2}_{+-} \rangle$ & $\langle r^{-2}_{--} \rangle$ & $\langle r^{-3}_{+-} \rangle_{R}$ & $\langle r^{-3}_{--} \rangle_R$ \\
      \hline 
  0.44866729948727416 & 0.057361735606920684 & -0.6464931313032396 & 0.02081865566718334 \\
       \hline
 $\langle {\bf n}_{he} \cdot {\bf n}_{he} \rangle$ &  $\langle {\bf r}_{he} \cdot {\bf r}_{ee} \rangle$ & $\langle r^{-3}_{+-} \rangle$ & $\langle r^{3}_{--} \rangle_{R}$ \\
      \hline
  -0.0135118419396675 & 31.982482810233879 &   &  \\
      \hline
 $\langle r^{4}_{+-} \rangle$ & $\langle r^{6}_{+-} \rangle$ & $\langle r^{8}_{+-} \rangle$ & $\langle r^{10}_{+-} \rangle$ \\
      \hline
  4.7926230075698435$\cdot 10^{3}$ & 1.7310611752766596$\cdot 10^{6}$ & 1.154800014201166$\cdot 10^{9}$ & 1.23351478796297$\cdot 10^{12}$ \\
      \hline
 $\langle r^{4}_{--} \rangle$ & $\langle r^{6}_{+-} \rangle$ & $\langle r^{8}_{+-} \rangle$ & $\langle r^{10}_{+-} \rangle$ \\
      \hline
  1.0465192805471107$\cdot 10^{4}$ & 3.7131795135786928$\cdot 10^{6}$ & 2.4329686306363651$\cdot 10^{9}$ & 2.57574765826309$\cdot 10^{12}$ \\
     \hline
 $\langle [r_{+-} r_{+-}]^{-1} \rangle$ & $\langle [r_{+-} r_{--}]^{-1} \rangle$ & $\langle [r_{32} r_{31} r_{21}]^{-1} \rangle$ & $\langle \delta({\bf r}_{+--}) \rangle$ \\
     \hline
 0.1446127143555099 & 0.0961529021408887 & 4.4982618400869338$\cdot 10^{-2}$ & 1.912535318$\cdot 10^{-4}$ \\
     \hline
 $\langle \frac{{\bf r}_{31} {\bf r}_{32}}{r_{31}^{-3}} \rangle$ & $\nabla_e \cdot \nabla_e$ & $\nu_{+-}^{(a)}$ & $\nu_{--}^{(a)}$ \\
     \hline
 -0.19196365724862741 & -1.303107187863320$\cdot 10^{-3}$ & -6.3636363629202 & -0.499999942324 \\
      \hline\hline
  \end{tabular}
  \end{center}
 ${}^{(a)}$The expected cusp values (in $a.u.$ for the $(1.75 m_e)^{+} e^{-} e^{-}$ ion are $\nu_{+-} = -0.636363636363636363636\ldots$ and $\nu_{--} = 0.5$ (exactly).
  \end{table}
\newpage
\begin{table}[tbp]
\caption{The coefficients $C_k$ in the mass-interpolation formulas determined for some bound state properties 
         in the two-electron $Z^{+} e^{-} e^{-}$ exitonic complexes (for the ground $1^{1}S-$states).}
     \begin{center}
     \scalebox{0.75}{%
     \begin{tabular}{| l | c | c | c |}
       \hline\hline 
 $C_k$ & $\langle r^{-1}_{+-} \rangle$ & $\langle r_{+-} \rangle$ & $\langle r^{+2}_{+-} \rangle$ \\ 
     \hline
 $C_0$    & 0.339821023059220307E+00 &  0.548963325235944993E+01 &  0.484189372262379554E+02 \\
 $C_1$    & 0.162512350434026144E+00 & -0.234594187716521358E+01 & -0.384174495144437727E+02 \\
 $C_2$    & 0.825685779319692414E-01 & -0.219316605264075911E+00 &  0.265101294063914059E+01 \\
 $C_3$    & 0.436659852608662058E-01 & -0.121229468881215481E+00 & -0.105461290243460516E+01 \\
 $C_4$    & 0.237954153106372529E-01 & -0.636771523924283964E-01 & -0.330076559557372832E+00 \\
 $C_5$    & 0.132544684410731888E-01 & -0.284408660570004830E-01 &  0.598647053486253223E-01 \\
            \hline 
 $C_6$    & 0.749782409001196623E-02 & -0.927905816490406999E-02 &  0.188318333178334562E+00 \\
 $C_7$    & 0.426021566550345151E-02 & -0.485079931818257154E-03 &  0.179718169057797372E+00 \\
 $C_8$    & 0.255890662698145735E-02 &  0.233623472236537070E-02 &  0.134576039470547131E+00 \\
 $C_9$    & 0.100102757003122304E-02 &  0.317285260352291693E-02 &  0.419842606050739852E-01 \\
 $C_{10}$ & 0.212318310050075383E-02 &  0.574649072626722250E-03 &  0.118236278837804872E+00 \\
          \hline
 $C_{11}$ & -0.263984988928883774E-02 &  0.453448002344733582E-02 & -0.187378742414278196E+00 \\
 $C_{12}$ &  0.588686513657035636E-02 & -0.537589805040799530E-02 &  0.322345121940872466E+00 \\
 $C_{13}$ & -0.723151430155999235E-02 &  0.809968011388563936E-02 & -0.396676657303459207E+00 \\
 $C_{14}$ &  0.690591033532986329E-02 & -0.742006803178840184E-02 &  0.314830775002427653E+00 \\
 $C_{15}$ & -0.391038375103147152E-02 &  0.451052046223482579E-02 & -0.163086987714568510E+00 \\
 $C_{16}$ &  0.118698098333386632E-02 & -0.151413858031999146E-02 &  0.331436669407528617E-01 \\
        \hline\hline
 $C_k$ & $E$ & $\langle r^{2}_{--} \rangle$ & $\langle \delta({\bf r}_{+-}) \rangle$ \\ 
     \hline
 $C_0$    & -0.262005070232980108E+00 &  0.931786338479813290E+02 &  0.207331980052196000E-01 \\
 $C_1$    & -0.128766481160936767E+00 & -0.687235705348342944E+02 &  0.306662376470630710E-01 \\
 $C_2$    & -0.651868609408879597E-01 &  0.232172844272751218E+01 &  0.304496129544827348E-01 \\
 $C_3$    & -0.335530172205839989E-01 & -0.214913007572575459E+01 &  0.258769197866279076E-01 \\
 $C_4$    & -0.175147829896120875E-01 & -0.684216329109166018E+00 &  0.836082416267927813E-02 \\
 $C_5$    & -0.927766709042202590E-02 &  0.104583108209125909E+00 &  0.158480690502420219E+00 \\
        \hline
 $C_6$    & -0.499854677870298553E-02 &  0.367509803811797095E+00 & -0.137728593152297410E+01 \\
 $C_7$    & -0.272620863234704363E-02 &  0.355915588152919500E+00 &  0.970476149406901520E+01 \\
 $C_8$    & -0.161603066046435153E-02 &  0.260099608747384438E+00 & -0.507385526411185295E+02 \\
 $C_9$    & -0.540623608818615660E-03 &  0.107315610658362776E+00 &  0.200702689865580268E+03 \\
 $C_{10}$ & -0.157460839398342092E-02 &  0.167317000591660025E+00 & -0.601492538670525686E+03 \\
       \hline
 $C_{11}$ &  0.225247088916263931E-02 & -0.229908519711612371E+00 &  0.135734707077270534E+04 \\
 $C_{12}$ & -0.480041885578094042E-02 &  0.417989841164496252E+00 & -0.226868662894445172E+04 \\
 $C_{13}$ &  0.598265472420348164E-02 & -0.538932981709009423E+00 &  0.272263096035267330E+04 \\
 $C_{14}$ & -0.566730968776260952E-02 &  0.435720638432029075E+00 & -0.221763692125528367E+04 \\
 $C_{15}$ &  0.320195965399203008E-02 & -0.236213354487059062E+00 &  0.109750767550857580E+04 \\
 $C_{16}$ & -0.957222950638015217E-03 &  0.471139717840663531E-01 & -0.249045057645281117E+03 \\
                     \hline\hline        
  \end{tabular}}
  \end{center} 
  \end{table}

\begin{thebibliography}{99}
\bibitem{XX} It is straightforward to show the completeness of the corresponding angular parts of the exponential basis functions
for the $S(L = 0)-$states considered in this study.
\end{thebibliography}
\end{document}